\documentstyle[12pt,epsf]{article}
\newcommand{\alt}{\,\rlap{\lower 3.5 pt \hbox{$\mathchar \sim$}} \raise 1pt
 \hbox {$<$}\,}

\begin{document}
\title{\vskip-3cm{\baselineskip14pt
\centerline{\normalsize MPI/PhT/97--076\hfill}
\centerline{\normalsize hep--ph/9711443\hfill}
\centerline{\normalsize Talk \# 502\hfill}
\centerline{\normalsize November 1997\hfill}
}
\vskip1.5cm
Colour-Octet Contributions to $J/\psi$ Production via Fragmentation at
HERA\thanks{To appear in {\it Proceedings of the International Europhysics
Conference on High Energy Physics}, Jerusalem, Israel, 19--26 August 1997.}}
\author{{\sc Bernd A. Kniehl}\\
Max-Planck-Institut f\"ur Physik (Werner-Heisenberg-Institut),\\
F\"ohringer Ring 6, 80805 Munich, Germany}
\date{}
\maketitle
\begin{abstract}
We study $J/\psi$ photoproduction via fragmentation at next-to-leading order
in the QCD-improved parton model, using the nonrelativistic factorization
formalism proposed by Bodwin, Braaten, and Lepage.
We predict that measurements of $J/\psi$ photoproduction at DESY HERA should
show a distinctive excess over the expectation based on the colour-singlet
model at small values of the inelasticity variable~$z$.
\end{abstract}

Since its discovery 25 years ago, the $J/\psi$ meson has provided a useful
laboratory for quantitatively testing quantum chromodynamics (QCD) and, in
particular, the interplay of perturbative and nonperturbative phenomena.
Recently, the cross section of $J/\psi$ inclusive production measured at the
Fermilab Tevatron turned out to be more than one order of magnitude in excess
of what used to be the best theoretical prediction, based on the
colour-singlet model (CSM).
As a solution to this puzzle, Bodwin, Braaten, and Lepage \cite{bod} proposed
the existence of so-called colour-octet processes to fill the gap.
The idea is that $c\bar c$ pairs are produced at short distances in
colour-octet states and subsequently evolve into physical (colour-singlet)
charmonia by the nonperturbative emission of soft gluons.
The underlying theoretical framework is provided by nonrelativistic QCD
(NRQCD) endowed with a particular factorization theorem, which implies a
separation of short-distance coefficients, which are amenable to perturbative
QCD, from long-distance matrix elements, which must be extracted from
experiment.
This formalism involves a double expansion in the strong coupling constant
$\alpha_s$ and the relative velocity $v$ of the bound $c$ quarks, and takes
the complete structure of the charmonium Fock space into account.

In order to convincingly establish the phenomenological significance of the
colour-octet mechanism, it is indispensable to identify it in other kinds of
high-energy experiments as well.
In this presentation, we briefly report on a next-to-leading-order (NLO)
analysis \cite{kni} of inelastic $J/\psi$ photoproduction via fragmentation at
HERA.  
We study direct and resolved photoproduction of prompt $J/\psi$ mesons and
$\chi_{cJ}$ mesons ($J=0,1,2$) that radiatively decay to $J/\psi+\gamma$,
taking into account the formation of both colour-singlet and colour-octet
$c\bar c$ states.
The dominant $J/\psi$ ($\chi_{cJ}$) Fock states are
$[\,\underline{1},{}^3\!S_1]$ and $[\,\underline{8},{}^3\!S_1]$
($[\,\underline{1},{}^3\!P_J]$ and $[\,\underline{8},{}^3\!S_1]$),
respectively.
For comparison, we also consider $\gamma g$ fusion in the CSM.

\begin{figure}[ht]
\begin{tabular}{cc}
\parbox{7.5cm}{\epsfxsize=7.5cm\epsffile[54  172  541  614]{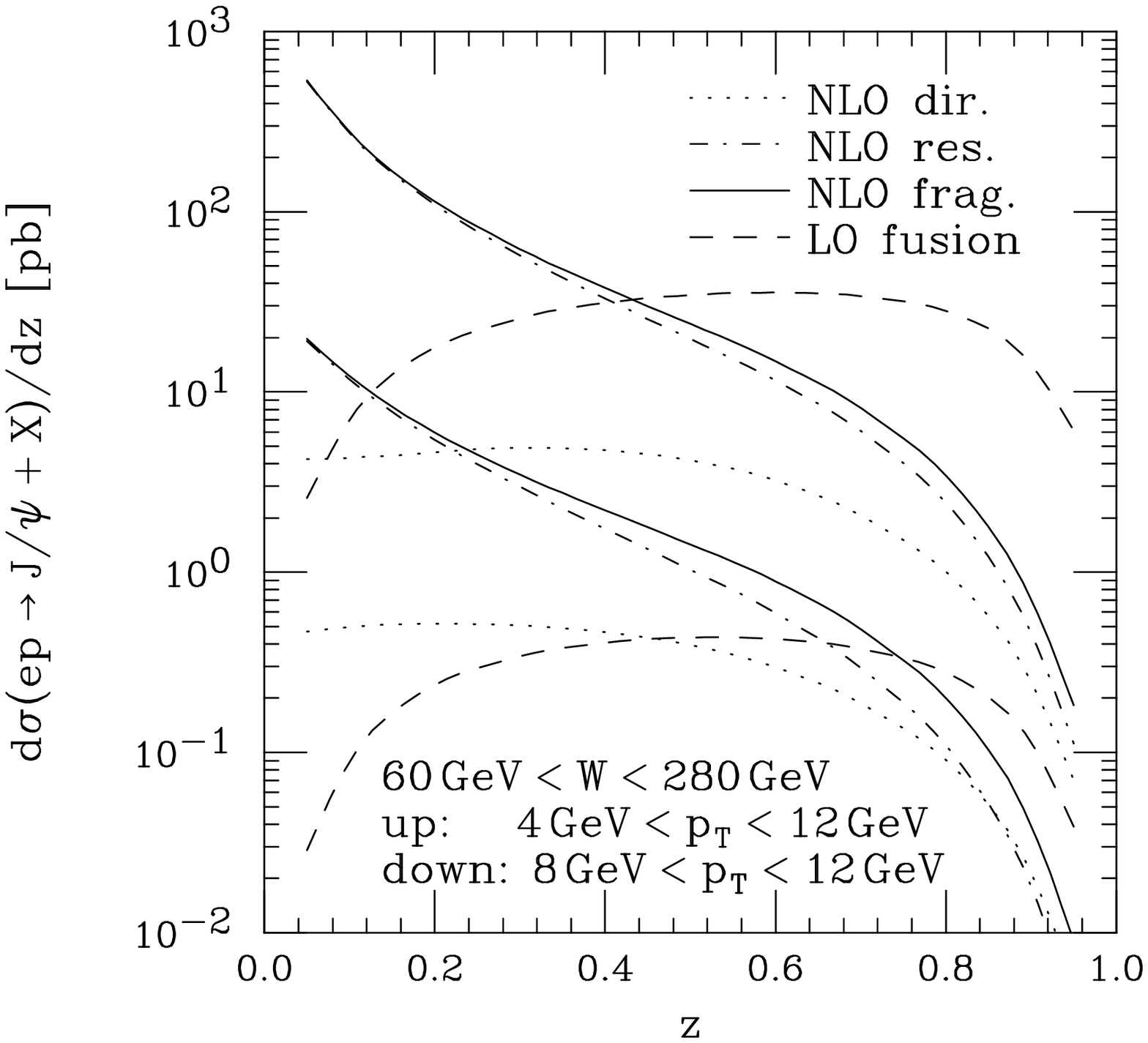}}&
\parbox{7.5cm}{\epsfxsize=7.5cm\epsffile[52  168  539  604]{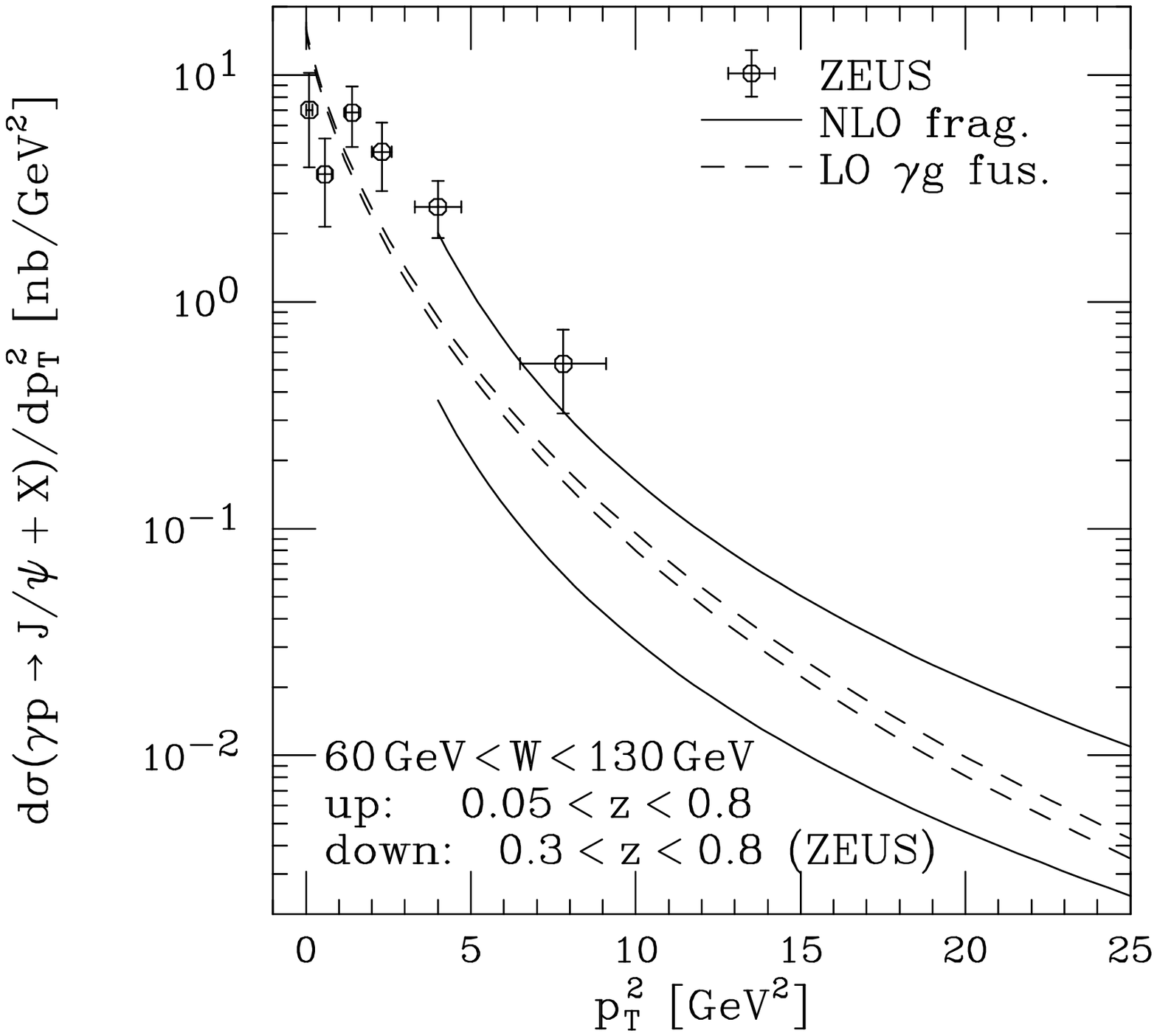}}
\end{tabular}
\caption{(a) $d\sigma/dz$ and (b) $d\sigma/dp_T^2$ for inelastic $J/\psi$
photoproduction at HERA.}
\label{f1}
\end{figure}

From Fig.~\ref{f1}a, we observe that the fragmentation mechanism vastly
dominates inelastic $J/\psi$ photoproduction in the lower range of the
inelasticity variable $z=p_p\cdot p_{J/\psi}/p_p\cdot p_\gamma$.
For a minimum-$p_T$ cut of 4~GeV (8~GeV), its contribution exceeds the one due
to $\gamma g$ fusion for $z<0.4$ (0.75), by factors of about 4 and 200 (20 and
700) at $z=0.25$ and 0.05, respectively.
The bulk of the fragmentation contribution is induced by resolved photons.
In Fig.~\ref{f1}b, the ZEUS $J/\psi\to\mu^+\mu^-$ data \cite{col} collected in
the interval $0.3<z<0.8$ are compared with the respective fragmentation and
$\gamma g$-fusion predictions.
We have divided the experimental data points by the estimated extrapolation
factor 1.07 which was included in Ref.~\cite{col} to account for the
unmeasured contribution from $0<z<0.3$.
In fact, our combined analysis of fragmentation and $\gamma g$ fusion suggests
that, at $p_T=5$~GeV, this factor should be as large as 3.4.
The corresponding value for $0.05<z<0.3$ is 2.6.

In conclusion, the cross section of inelastic $J/\psi$ photoproduction in $ep$
collisions at low to intermediate $z$ and large $p_T$, where fragmentation
production is dominant, is very sensitive to the colour-octet matrix element
$\langle0|{\cal O}^{J/\psi}[\,\underline{8},{}^3\!S_1]|0\rangle$.
By contrast, $\gamma g$ fusion, which is only relevant in the upper $z$ range
and for $p_T\alt M_{J/\psi}$, probes
$\langle0|{\cal O}^{J/\psi}[\,\underline{8},{}^1\!S_0]|0\rangle$ and
$\langle0|{\cal O}^{J/\psi}[\,\underline{8},{}^3\!P_J]|0\rangle$.
We propose to accordingly extend previous measurements of this cross section
at HERA, in order to obtain an independent, nontrivial check of the Tevatron
colour-octet charmonium puzzle.

\smallskip

{\it Acknowledgements.} I would like to thank Gustav Kramer for his
collaboration on the work presented here.


\begin{thebibliography}{99}
\bibitem{bod} G.T. Bodwin, E. Braaten, and G.P. Lepage,
Phys.\ Rev.\ D 51 (1995) 1125; 55 (1997) 5855 (E).
\bibitem{kni} B.A. Kniehl and G. Kramer,
Phys.\ Rev.\ D 56 (1997) 5820;
Phys.\ Lett.\ B 413 (1997) 416.
\bibitem{col} ZEUS Collaboration, J. Breitweg et al.,
Contributed Paper No.\ pa02--047 to the 28th International Conference on High
Energy Physics, Warsaw, Poland, 25--31 July 1996.
\end{thebibliography}
\end{document}